\documentclass[aps]{revtex4}
\usepackage{epsfig}
\usepackage{graphicx}
\usepackage{dcolumn}
\usepackage{amsmath}

\begin{document}

\title{Neutron Transfer Studied with a Radioactive beam of $^{24}$Ne, using TIARA at SPIRAL}

\author{W.~N. CATFORD$^1$, C.~N. TIMIS$^1$, R.~C. LEMMON$^2$, M. LABICHE$^3$, N.~A. ORR$^4$, L. CABALLERO$^5$,
R. CHAPMAN$^3$, M. CHARTIER$^6$,
M. REJMUND$^7$, H. SAVAJOLS$^7$ and the TIARA COLLABORATION}

\address{
$^1$Department of Physics, University of Surrey, Guildford, Surrey GU2 7XH, UK \\
$^2$CCLRC Daresbury Laboratory, Daresbury, Warrington WA4 4AD, UK \\
$^3$University of Paisley, Paisley, Scotland PA1 2BE, UK      \\
$^4$LPC, IN2P3-CNRS, ISMRA and Universit\'e de Caen, F-14050 Caen, France \\
$^5$IFC, CSIC-Universidad de Valencia, E-46071 Valencia, Spain \\
$^6$Department of Physics,  The University of Liverpool, Liverpool L69 7ZE, UK \\
$^7$GANIL, BP 55027, 14076 Caen Cedex 5, France \\
E-mail: W.Catford@surrey.ac.uk
}

\begin{abstract}
A general experimental technique for high resolution studies of nucleon transfer reactions
using radioactive beams is briefly described, together with the first new physics results
that have been obtained with the new TIARA array. These first results from TIARA are for
the reaction $^{24}$Ne(d,p)$^{25}$Ne, studied in inverse kinematics with a pure radioactive
beam of $10^5$ pps from the SPIRAL facility at GANIL. The reaction probes the energies of neutron
orbitals relevant to very neutron rich nuclei in this mass region and the results highlight the
emergence of the N=16 magic number for neutrons and the associated disappearance of the
N=20 neutron magic number for the very neutron rich neon isotopes.\end{abstract}

\maketitle

\section{Introduction}\label{sec:Intro}

A great hope for the future in radioactive beam experiments is to be able to map
out the changing shell structure for very exotic nuclei, away from stability,
where this arises from effects such as the monopole migration of orbital energies
and the changes brought about by alterations in the surface environment and
spin-orbit splitting\cite{grawe}.
Nucleon transfer reactions such as (d,p), (p,d) etc.\   are
an established means of populating and studying nuclear levels that have a
substantial single-particle structure. The development of techniques to use
such reactions with radioactive beams, across a wide range of beam energies and
masses and with high energy resolution, will open the way to exploit transfer
across new regions of the nuclear chart and hence to study the new nuclear
structure effects that evolve.

The technique that is described here, and implemented via the new TIARA array
used in association with the VAMOS spectrometer and the EXOGAM gamma-ray array,
is designed to achieve excitation energy resolution of better than 20-40 keV in
the final nucleus. This is an order of magnitude better than can be achieved in
a reasonable experimental setup that uses charged-particle observations
only\cite{winfield}.
The complete kinematical detection of the binary transfer reaction products
specifies the reaction channel cleanly, where the identification of the heavy
(beam-like) particle at least in Z is required, and the light (target-like)-particle
detection allows angular distributions to be measured for any mass of projectile.

The present paper updates and extends results of the analysis in progress, reported
elsewhere\cite{nustar,enam}.

\section{The TIARA Array}

The requirement to use inverse kinematics in order to study nucleon transfer
reactions, induced on radioactive species by protons and deuterons, imposes
certain rather general requirements on the detection system to be used. The
kinematics turns out to lead to particular reactions always appearing in the
same characteristic range of laboratory angles and with similar energies,
regardless of the mass or velocity of the incident beam\cite{rnb,enam}. This
allows a general purpose transfer apparatus to be designed.

\begin{figure}[ht]
\centerline{\psfig{figure=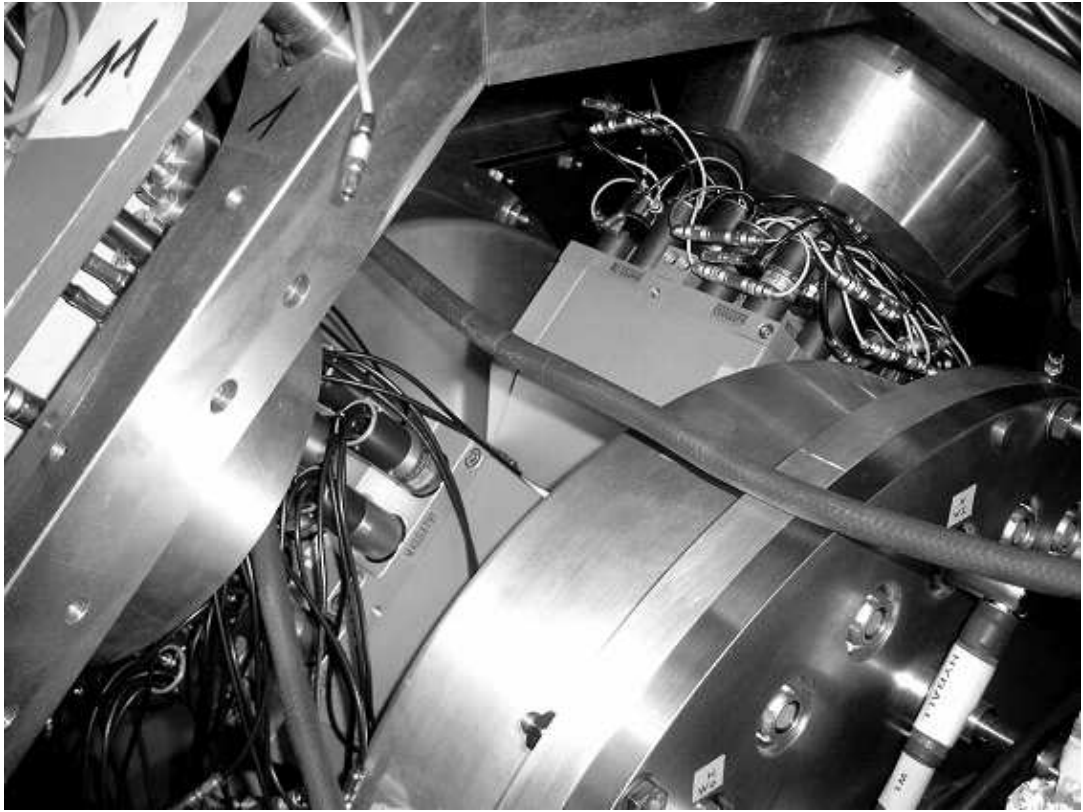,width=0.8\columnwidth,angle=0}}

\caption{The four EXOGAM detectors, in a compact cube geometry, are shown
mounted around the TIARA vacuum vessel, which is symmetric (both cylindrically
and forward-backward around the target) and narrows down to just 98mm in diameter near
the target. The TIARA array is inside, and the beam enters through a target
selection mechanism, also located inside the vessel at the lower right.\label{TIARA-pic}}
\end{figure}

The design philosophy and detailed description
of TIARA has been discussed elsewhere\cite{nustar,denton}
but briefly the aim was to surround the target with a charged particle array
that approached $4\pi$ coverage, with reasonable energy measurment and an angular
resolution of 1 or 2 degrees. This array needed to be very compact so that a
high gamma-ray efficiency of $>$15\% (at 1.3 MeV) could be achieved, whilst
avoiding the exposure of gamma-ray detectors to decay radiation from  beam
particles scattered in the forward 40$^\circ$. In the present setup, the TIARA
array covers 82\% of $4\pi$ with active silicon and the gamma-ray detectors
are in a close cube geometry and thus subtend 67\% of $4\pi$. The setup is mounted
in front of a magnetic spectrometer which is used to separate physically the direct
beam and the transfer reaction products, after the target. The region around the target
is shown in fig. \ref{TIARA-pic}.

\section{Experimental Details}

The TIARA system was set up in front of the VAMOS spectrometer at GANIL\cite{vamos},
which was operated in dispersive mode at zero degrees. Direct beam was intercepted
just in front of the focal plane detectors. The support frame and four detectors of
EXOGAM\cite{exogam} surrounded
the TIARA chamber. 
All events in which a particle was detected in TIARA were recorded. The gamma ray
parameters were recorded via the VAMOS acquisition system and events were correlated
with TIARA in real time via an event stamping method developed at GANIL.

%

An isotopically pure beam of $^{24}$Ne was supplied at 10 MeV/nucleon after
reacceleration in the CIME cyclotron connected to the SPIRAL facility at GANIL.
The beam intensity of $10^5$ pps was a factor of two
lower than the maximum due to a limitation placed on the emittance, which limited the
beam spot on target to a diameter of approximately 2mm base width. The target was
1 mg/cm$^2$ of CD$_2$ self supporting on a thin 25mm diameter frame.

A test experiment was performed with a stable beam of $^{14}$N at similar intensity
and beam quality, in order to verify that normal kinematics (d,p) results from the
literature could be reproduced with the TIARA setup.
Good agreement was found\cite{labiche}.

\section{Results}

\begin{figure}[ht]
\centerline{\epsfxsize=4.0in\epsfbox{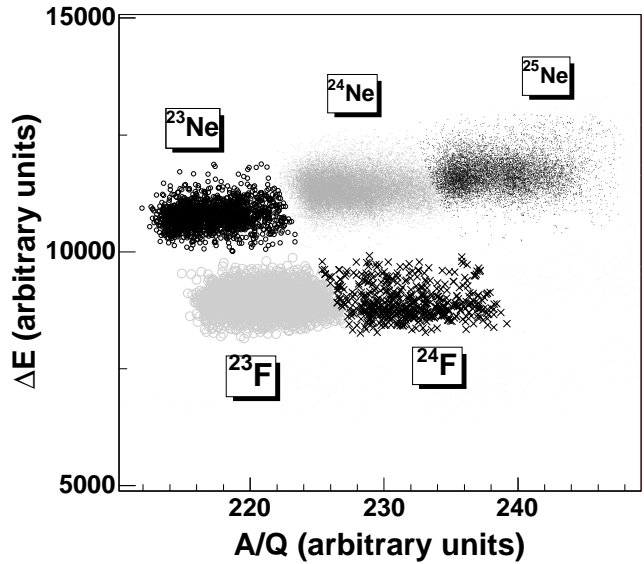}}

\caption{Particle identification for particles recorded behind the beam stop
at the focal plane of VAMOS.\label{PID}}
\end{figure}

The isotopic identification for beam-like particles recorded at the focal plane of
VAMOS is shown in fig.\ref{PID}. This is derived from measured $\Delta E$, $E$
and time-of-flight parameters plus a ray tracing calculation that used the
horizontal and vertical angles and positions measured at the focal plane. The
ray tracing algorithm employed a neural network that was trained using a set of theoretically
calculated rays obtained by numerical integration of their trajectories through
VAMOS\cite{cozmin} and this gives results identical to an algebraic algorithm developed
at GANIL.

\begin{figure}[ht]
\centerline{\epsfxsize=4.1in\epsfbox{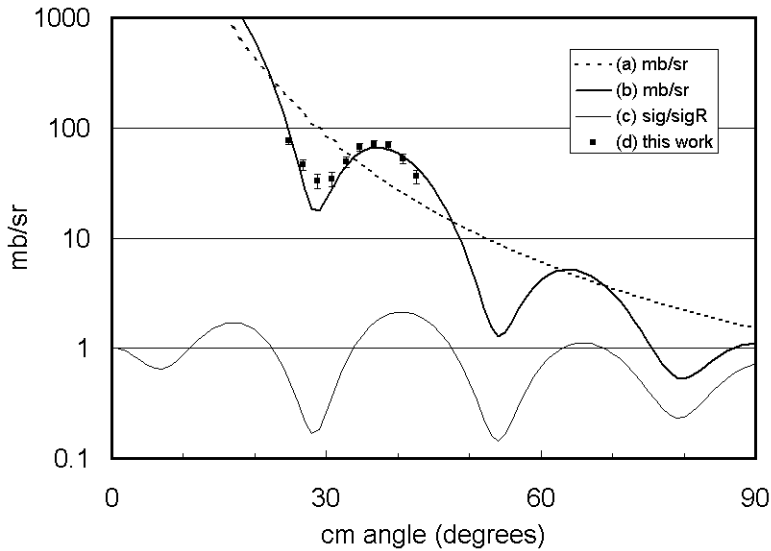}}

\caption{Angular distribution for d + $^{24}$Ne elastic scattering:
(a) Rutherford formula, (b) optical model calculation (see text), (c) ratio of
(b) to (a), and (d) present measurements. \label{elastics}}
\end{figure}

By selecting the $^{24}$Ne ions in fig. \ref{PID}, the scattered deuterons recorded
in TIARA could be analysed. The $^{24}$Ne momentum changes sufficiently quickly with
scattering angle that very forward scattered elastics can still avoid the beam stop.
The energy of the deuterons changes rapidly with their angle\cite{rnb} and, by using
energy cuts, the elastic angular distribution could be extracted (see fig. \ref{elastics}).
A good fit was obtained using the optical potential measured for d+$^{26}$Mg at a
similar energy\cite{meurders}.
The normalisation obtained using these elastic data allowed absolute transfer cross sections
to be extracted with confidence.

\begin{figure}[ht]
\centerline{\epsfxsize=4.1in\epsfbox{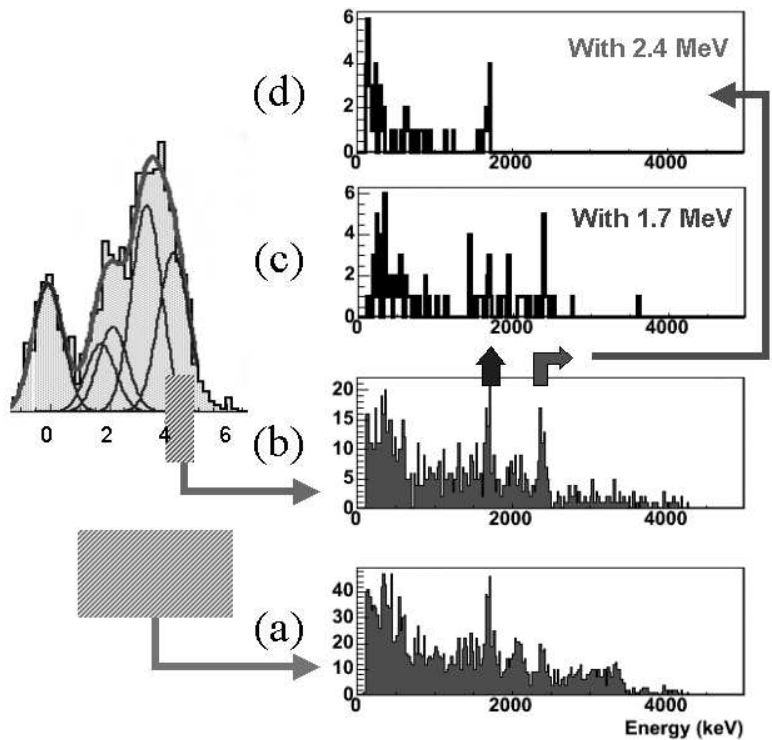}}

\caption{Gamma ray energy spectra from d($^{24}$Ne,p$\gamma$)$^{25}$Ne gated according to
the excitation energy spectrum derived from the proton energy and angle (shown
in inset): (a) all excited states, (b) peak near 4 MeV, (c) as for (b) but also requiring
a coincident 1.7 MeV gamma ray, (d) as for (b) but requiring a 2.4 MeV gamma ray. \label{gammas}}
\end{figure}

From the measured energy and angle recorded for protons from the (d,p) reaction to
make $^{25}$Ne (cf. ref. \cite{nustar}) the excitation energy spectrum for states
in $^{25}$Ne could be deduced. Different peaks in this spectrum could be used to
gate the spectrum of associated gamma rays. Example data are included in fig, \ref{gammas}.
An important result of this analysis was that the excitation energies of the
populated states could be fixed with an accuracy of order 30 keV. The limiting factor
in this accuracy was the poor statistics of the gamma ray spectrum. This was in fact
severely compromised in the present experiment by an intermittent fault in an electronic
discriminator unit, and the eventual aim in this type of experiment will be to use individual
gamma ray peaks to apply gates in the analysis. In the present case, however, it was
still of vital importance that the gamma ray data could fix the energies and the number
of peaks to be fitted to the (poorer resolution) excitation energy spectra derived
from the particle energies. These fits are shown in the inset of fig. \ref{gammas}.
The data are just sufficient to allow limited gamma-gamma coincidence analysis. In the case
of the state near 4 MeV it can be seen that the 1.7 MeV and 2.4 MeV gamma rays seen in
its decay (fig. \ref{gammas}(b)) appear to be in coincidence (fig. \ref{gammas} (c)
and (d)).

The excitation energy spectrum (derived from the proton energy and angle) was fitted with
5 peaks. The widths of these peaks depended on the experimental resolution of the system, and this
was in turn dependent on the detection angle of the proton. Thus, the data were binned for
angular regions chosen to be 8$^\circ$ wide in order to give sufficient statistics for fitting.
In order to fit the angular distributions, different optical model potentials were
investigated, taken from (d,p) reactions on neighbouring nuclei $^{26}$Mg\cite{meurders}
and $^{22}$Ne\cite{howard}. A systematic comparison with adiabatic calculations according
to the prescription of Johnson and Soper\cite{johnson} was also performed. The adiabatic
analysis was adopted for the extraction of spectroscopic factors, which were determined by
normalising the theoretical curve to the data for each state, with particular emphasis
placed on the data for the smallest center of mass angles
(closest to 180$^\circ$ in the laboratory)\cite{tsang}.

The results of the analysis are included in Table \ref{specfacs}. The identifications of
the spins are discussed below. In general, spectroscopic factors extracted in this fashion
have an uncertainty of order 20\% arising from the assumptions in the reaction theory,
and this is the dominant source of uncertainty in the quoted results.

\begin{table}[ph]
\caption{Results for states in $^{25}$Ne identified as being populated in neutron transfer
on $^{24}$Ne. Previous E$_{\rm x}$ is from
Reed {\em et al.} and USD refers to a
$1s0d$ shell model calculation.}
{\footnotesize
\begin{tabular}{ccccccc}
\hline
{} &{} &{} &{} &{}\\[-1.5ex]
E$_{\rm x}$ (MeV) & E$_{\rm x}$ (MeV) & $\ell$ (~$\hbar$~) & $J^\pi$ & $S$ & $S$ & E$_{\rm x}$ (MeV) \\
present & previous & transfer & & present & USD & USD \\[1ex]
\hline
{} &{} &{} &{} &{} &{}\\[-1.5ex]
0 & 0  & 0 & 1/2$^+$ & 0.80 & 0.63 & 0\\[1ex]
1.680 & 1.703  & 2 & 5/2$^+$ & 0.15 & 0.10 & 1.779 \\[1ex]
2.030 & -  & 2 & 3/2$^+$ & 0.44 & 0.49 & 1.687 \\[1ex]
3.330 & -  & 1 & 3/2$^-$ & 0.75 & - & -\\[1ex]
4.030 & -  & (3) & 7/2$^-$ & 0.73 & - & -\\[1ex]
\hline
\end{tabular}\label{specfacs} \\
}
\vspace*{-13pt}
\end{table}

\section{Discussion}

The key feature emerging from Table \ref{specfacs} is that the state identified as
the first 3/2$^+$ state in $^{25}$Ne, which reflects most directly the single
particle energy of the $0d_{3/2}$ shell model orbital, is substantially higher
than predicted. The identification rests on both the relative strength of this ``particle''
state compared to the $0d_{5/2}$ ``hole'' state and the observed gamma decay
pathways. The shift of order 350 keV is presumably due to matrix elements
in the USD effective interaction that are not well determined from data on less
neutron rich nuclei. The shift can be understood very naturally in the monopole
shift picture\cite{otsuka,grawe}, in which the emptying of the $d_{5/2}$ proton orbital
in the more neutron rich N=15 isotones removes an attractive interaction that
lowers the neutron $0d_{3/2}$ energy for nuclei closer to stability. This tends to make
N=16 a magic number for neutron rich nuclei. Simultaneously, the
gap to the negative parity orbitals $0f_{7/2}$ and $1p_{3/2}$ is reduced
and N=20 loses its magicity\cite{utsuno}.

The state identified as the 5/2$^+$ is almost certainly the state seen in
beta decay\cite{reed} at 1.703 MeV and has also been seen recently
in neutron knockout from $^{26}$Ne\cite{terry}. This latter observation also
supports the identification of the 1.703 MeV level as the hole state and the newly observed
level at 2.03 MeV as the 3/2$^+$ state. The further implications of these results
are still under investigation.

\section*{Acknowledgments}
We acknowledge with thanks the support of the GANIL and LPC Caen technicians,
during the installation and commissioning of the TIARA array. Mr Geoffrey Moores (University
of Paisley) and the Daresbury design staff are thanked for their vital contributions.
This work was supported in the UK by EPSRC grants held at Surrey, Paisley, Daresbury and Birmingham.

\end{document}